\documentclass[11pt,fleqn]{article}
\usepackage{amssymb}

\usepackage{amsmath}

\usepackage{amsmath,amssymb,amsthm}

\begin{document}

\begin{center}
{\Large \textbf{Second-Order Differential Invariants of the Rotation Group  \\[0pt]
$\bf{O(n)}$ and of its Extensions: $\bf{E(n)}$, $\bf{P(1,n)}$,
$\bf{G(1,n)}$ }}\footnote{Acta Appl. Math., 1992, {\bf 28},
69--92; some misprints were corrected.}

\vskip 3pt {\large \textbf{W.I. FUSHCHYCH, Irina YEHORCHENKO }}
\vskip 6pt Institute of
Mathematics of NAS of Ukraine, 3 Tereshchenkivska Str., Kyiv 4, Ukraine\\[0pt%
]
E-mail: iyegorch@imath.kiev.ua
\end{center}

\begin{abstract}
\noindent Functional bases of second-order differential invariants
of the Euclid, Poincar\'e, Galilei, conformal, and projective
algebras are constructed. The results obtained allow us to
describe new classes of nonlinear many-dimensional invariant
equations.
\end{abstract}

\centerline{\bf Introduction}

The concept of the invariant is widely used in various domains of
mathematics. In this paper, we investigate the differential
invariants within the framework of symmetry analysis of
differential equations.

Differential invariants and construction of invariant equations
were considered by S.~Lie [1] and his followers [2, 3]. Tresse [2]
had proved the theorem on the existence and finiteness of a
functional basis of differential invariants. However, there exist
quite a few papers devoted to the construction in explicit form of
differential invariants for specific groups involved in mechanics
and mathematical physics.

Knowledge of differential invariants of a certain algebra or group
facilitates clas\-si\-fi\-ca\-tion of equations invariant with
respect to this algebra or group. There are also some general
methods for the investigation of differential equations which need
tide explicit form of differential invariants for these equations'
symmetry groups (see e.g. [3, 4]).

A brief review of our investigation of second-order differential
invariants for the Poincar\'e and Galilei groups is given in [5,
6]. Our results on functional bases of differential invariants are
founded on the Lemma about functionally independent warrants for
the proper orthogonal group and two $n$-dimensional symmetric
tensors of the order 2.

We should like to stress that we consider functionally independent
invariants of but not irreducible ones, as in the classical theory
of invariants.

Bases of irreducible invariants for the group $O(3)$ and
three-dimensional symmetric tensors and vectors are adduced in
[7].

The definitions of differential invariants differ in various
domains of mathematics, e.g. in differential geometry and symmetry
analysis of differential equations. Thus, we believe that some
preliminary notes are necessary, though these formulae and
definitions can be found in [8, 9, 10].

We deal with Lie algebras consisting of the infinitesimal
operators

\renewcommand{\theequation}{0.\arabic{equation}}
\setcounter{equation}{0}
\begin{equation}
X=\xi^i(x,u)\partial_{x_i}+\eta^r(x,u)\partial_{u^r}.
\end{equation}

Here $x=(x^1,x^2,\ldots,x^n)$, $u=(u^1,\ldots,u^m)$. We usually
mean the summation over the repeating indices.

\smallskip

\noindent {\bf Definition 1.} {\it  The function \[ F = F(x,
u,\mathop{u}\limits_1,\ldots,\mathop{u}\limits_l),
\]
\noindent where $\mathop{u}\limits_k$ is the set of all
$k$th-order partial derivatives of the function $u$ is called
a~differential invariant for the Lie algebra $L$ with basis
elements $X_i$ of the form (0.1) $(L=\langle X_i\rangle)$ if it is
an invariant of the $l$-th prolongation of this algebra:
\begin{equation}
\mathop{X}\limits^l\!{}_s F(x,
u,\mathop{u}\limits_1,\ldots,\mathop{u}\limits_l)= \lambda_s(x,
u,\mathop{u}\limits_1,\ldots,\mathop{u}\limits_l)F, \end{equation}
\noindent where the $\lambda_s$ are some functions; when
$\lambda_i=0$, $F$ is called an absolute invariant; when
$\lambda_i \ne 0$, it is a relative invariant.}

\smallskip

Further, we deal mostly with absolute differential invariants and
when writing `differential invariant' we mean `absolute
differential invariant'.

\smallskip

\noindent {\bf Definition 2.} {\it A maximal set of functionally
independent invariants of order $r\leq l$ of the Lie algebra $L$
is called a functional basis of the $l$th-order differential
invariants for the algebra $L$.}

\smallskip

We consider invariants of order 1 and 2 and need the first and
second prolongations of the operator $X$ (0.1) (see, e.g.,
[8--11])
\[
\mathop{X}\limits^1=X+\eta_i^r\partial_{u_i^r}, \quad
\mathop{X}\limits^2=\mathop{X}\limits^1+\eta_{ij}^r\partial_{u_{ij}^r}
\]
\noindent the coefficients $\eta_i^r$ and $\eta_{ij}^r$ taking the
form

\[ \begin{array} {l}
\eta_i^r=(\partial_{x_i}+u_i^s\partial_{u^s})\eta^r-u_k^r(\partial_{x_i}+u_i^s\partial_{u^s})\xi^k,
\vspace{1mm} \nonumber \\
\eta_{ij}^r=(\partial_{x_i}+u_j^s\partial_{u^s}+u_{jk}^s\partial_{u_k^s})\eta_i^r
-u_{ik}^r(\partial_{x_j}+u_j^s\partial_{u^s})\xi^k. \nonumber
\end{array}
\]

While writing out lists of invariants, we shall use the following
designations

\begin{equation} \begin{array}{l}
u_a\equiv\frac{\partial u}{\partial x_a},\quad
u_{ab}\equiv\frac{\partial^2 u}{\partial x_a
\partial x_b}, \vspace{1mm}\\ S_k(u_{ab})\equiv
u_{a_1a_2}u_{a_2a_3}\cdots u_{a_{k-1}a_k}u_{a_ka_1},
\vspace{1mm}\\
S_{jk}(u_{ab},v_{ab})\equiv u_{a_1a_2}\cdots u_{a_{j-1}a_j}
v_{a_ja_{j+1}}\cdots v_{a_k a_1},\vspace{1mm}\\
R_k(u_a,u_{ab})\equiv u_{a_1}u_{a_k}u_{a_1a_2}u_{a_2a_3}\cdots
u_{a_{k-1}a_k}u_{a_ka_1}.
\end{array}
\end{equation}

Here and further we mean summation over the repeated indices from
1 to $n$. In all the lists of invariants, $k$ takes on the values
from 1 to $n$ and $j$ takes the values from 0 to $k$. We shall not
discern the upper and lower indices with respect to summation: for
all Latin indices
\[
x_ax_a\equiv x_a x^a\equiv x^a x_a=x_1^2+x_2^2+\cdots+x_n^2.
\]

\centerline{\bf 1. Differential invariants for the Euclid algebra}

The Euclid algebra $AE(n)$ is defined by basis operators
\renewcommand{\theequation}{1.\arabic{equation}}
\setcounter{equation}{0} \begin{equation} \partial_a\equiv
\frac{\partial}{\partial_{x_a}},\quad
J_{ab}=x_a\partial_b-x_b\partial_a.
\end{equation}

Here and further, the letters $a$, $b$, $c$, $d$, when used as
indices, take on the values from 1 to $n$, $n$ being the number of
space variables $(n\geq 3)$.

The algebra $AE(n)$ is an invariance algebra for a wide class of
many-dimensional scalar equations involved in mathematical physics
--- the Schr\"odinger, heat, d'Alem\-bert equations, etc.

In this section, we shall explain in detail how to construct a
functional basis of the second-order differential invariants for
the algebra $AE(n)$. This basis will be further used to find
invariant bases for various algebras containing the Euclid algebra
as a~subalgebra --- the Poincar\'e, Galilei, conformal, projective
algebras, etc.

\smallskip

{\bf 1.1. The main results.} Let us first formulate the main
results of the section in the form of theorems.

\smallskip

\noindent {\bf Theorem 1.} {\it There is a functional basis of
second-order differential invariants for the Euclid algebra
$AE(n)$ with the basis operators (1.1) for the scalar function
$u=u(x_1,\ldots,x_n)$ consisting of these $2n + 1$ invariants}
\begin{equation}
u,\quad S_k(u_{ab}),\quad R_k(u_a,u_{ab}).
\end{equation}

\noindent {\bf Theorem 2.} {\it The second-order differential
invariants of the algebra $AE(n)$ (1.1) for the set of scalar
functions $u^r$, $r = 1,\ldots,m$, can be represented as functions
of the following expressions:}
\begin{equation}
u^r,\quad S_{jk}(u^1_{ab},u_{ab}^r),\quad R_k(u_a^r,u_{ab}^1).
\end{equation}

{\bf 1.2.  Proofs of the theorems.} Absolute differential
invariants are obtained as solutions of a linear system of
first-order partial differential equations (PDE). Thus, the number
of elements of a functional basis is equal to the number of
independent integrals of this system. This number is equal to the
difference between the number of variables on which the functions
being sought depend, and the rank of the cor\-res\-pon\-ding
system of PDE (in our case, this rank is equal to the generic rank
of the prolonged operator algebra [8, 9].

To prove the fact that $N$ invariants which have been found,
$F^i=F^i(x,u,\mathop{u}\limits_1,\ldots, \mathop{u}\limits_l)$,
form a functional basis, it is necessary and sufficient to prove
the following sta\-te\-ments:
\begin{enumerate} \itemsep0mm
\item[(1)] the $F^i$ are invariants; \item[(2)] the $F^i$ are
functionally independent; \item[(3)] the set of invariants $F^i$
is complete or $N$ is equal to the difference of the number of
variables $(x,u,\mathop{u}\limits_1,\ldots, \mathop{u}\limits_l)$
 and the rank of the system of defining operators.
 \end{enumerate}

We seek second-order differential invariants in the form
\[
F=F(x,u,\mathop{u}\limits_1,\mathop{u}\limits_2).
\]

It follows from the condition of invariance with respect to
translation opera\-tors~$\partial_a$ that $F$ does not depend on
$x_a$; evidently, $u$ is an invariant of the operators (1.1).
Thus, it is sufficient to seek invariants depending on
$\mathop{u}\limits_1$ and $\mathop{u}\limits_2$ only. The
criterion of the absolute invariance (0.1) in this case has the
form
\begin{equation}
\hat J_{ab}F(\mathop{u}\limits_1,\mathop{u}\limits_2)=0,
\end{equation}
\noindent where
\begin{equation}
\hat J_{ab}=u_a^r\partial_{u_b^r}-u_b^r\partial_{u_a^r}+
2(u_{ac}^r\partial_{u_{bc}^r}-u_{bc}^r\partial_{u_{ac}^r}),
\end{equation}
\noindent the summation over $r$ from 1 to $m$ being implied.

In that way, the problem of finding the second-order differential
invariants of the algebra $AE(n)$ is reduced to the construction
of a functional basis for the rotational algebra $AO(n)$ with the
basis operators (1.5) for $m$ vectors and $m$ symmetric tensors of
order 2.

\smallskip

\noindent {\bf Lemma 1.} {\it  The rank of the algebra $AO(n)$ is
equal to $(n(n - 1))/2$.}

\smallskip

\noindent {\bf Proof.} It is sufficient to prove the lemma for $m
= 1$. The basis of the algebra (1.5) consists of $(n(n - l))/2$
operators. According to definition [8], its rank is equal to the
generic rank of the coefficient matrix of these operators. Let us
put $u_{ab}=0$ when $a\not= b$ and write down the coefficient
columns by $\partial_{u_{ab}}$ of the operators (1.5):
\begin{equation}
\left(\begin{array}{cccc}
u_{11}-u_{22} & 0 & \cdots & 0 \\
0 & u_{11}-u_{33} & \cdots & 0\\
\cdots & \cdots & \cdots & \cdots\\
0 & 0 & \cdots & u_{n-1,n-1}-u_{nn}\end{array}\right).
\end{equation}

When $u_{aa}\not= u_{bb}$ for $a\not= b$ and all $u_{aa}\not=0$,
the determinant of the matrix (1.6) does not vanish, therefore its
generic rank (that is, the generic rank the algebra being
considered) cannot be less than $(n(n - 1))/2$. The lemma is
proved. \hfill \rule{2mm}{2mm}

\smallskip

\noindent {\bf Lemma 2.} {\it  The expressions
\begin{equation}
S_k(u_{ab}), \quad R_k(u_a,u_{ab})
\end{equation}
\noindent are functionally independent.}

\smallskip

\noindent {\bf Proof.} To establish independence of expressions
(1.7), it is sufficient to consider the case when $u_{ab}=0$ if
$a\not=b$ and $u_{aa}\not=0$. Let us write down the Jacobian of
the invariants
\begin{equation}
\left| \begin{array}{ccc|ccc}
1 & \cdots & 1 & &&\\
2u_{11} & \cdots & 2u_{nn} & & \bf{0} &\\
\cdots & \cdots & \cdots &&\\
nu_{11}^{n-1} & \cdots & nu_{rr}^{n-1} & &&\\
\hline
&&& 2u_1 & \cdots & 2u_n\\
&\cdots && \cdots & \cdots & \cdots\\
&&& 2u_1 u_{11}^{n-1} & \cdots & 2u_n u_{nn}^{n-1}
\end{array}\, \right|
\end{equation}

The Jacobian (1.8) is equal up to a coefficient to the product of
two Vandermonde determinants and is not equal to zero if
$u_{aa}\not=u_{bb}$
 whenever $a\not=b$. Thus, the expressions (1.7) are
functionally independent.\hfill\rule{2mm}{2mm}

\smallskip

\noindent {\bf Proof of Theorem 1.} The fact that expressions
(1.2) are invariants of $AO(n)$ can be easily proved by direct
substitution of these expressions into the invariance conditions.
Nevertheless, it is useful to note that $S_k(u_{ab})$ are traces
of the symmetric matrix $(u_{ab})=U$ and its powers,
$R_k(u_a,u_{ab})$ are the scalar products of the vector
$(u_a)=(u_1,\ldots,u_n)$, the matrix $U^{k-1}$ and the vector
$(u_a)^T$.

The invariants for the vector $(u_a)$ and the symmetric tensor
$(u_{ab})$ depend on their $(n(n + 3))/2$ elements. Thus, it
follows from Lemma~1 that a functional basis of the algebra
$AO(n)$ for $(u_a)$ and $(u_{ab})$ must consist of
\[
\frac{n(n+3)}{2}-\frac{n(n-1)}{2}=2n
\]
\noindent invariants.

Therefore the set (1.7) is a complete set of functionally
independent invariants of the form $F =
F(\mathop{u}\limits_1,\mathop{u}\limits_2)$ and (1.2) represents a
functional basis of the second-order invariants for the algebra
$AE(n)$. The theorem is proved. \hfill\rule{2mm}{2mm}

\smallskip

Let us consider the case of two vectors $(u_a)$, $(v_a)$ and two
symmetric tensors of the second order $(u_{ab})$, $(v_{ab})$. The
operators of the rotation algebra have the form (1.5), $u\equiv
u^1$, $v\equiv u^2$.

In this case, a functional basis of invariants contains
\[
2\left(\frac{n(n-1)}{2}+2n\right)-\frac{n(n-1)}{2}=\frac{n(n+7)}{2}
\]
\noindent elements for which we take the following expressions
\begin{equation}
R_k(u_a,u_{ab}),\quad R_k(v_a,u_{ab}),\quad S_{jk}(u_{ab},v_{ab}).
\end{equation}

The invariance of expressions (1.9) with respect to the operators
(1.5) can be easily proved by their direct substitution to (1.4).
To establish their functional independence, we shall use the
following lemma.

\smallskip

\noindent {\bf Lemma 3.} {\it  Let
\[
U=(u_{ab})_{a,b=1,\ldots,n}, \quad V=(v_{ab})_{a,b=1,\ldots,n}
\]
\noindent be symmetric matrices. Then the expressions
\begin{equation}
S_{jk}(u_{ab},v_{ab})={\rm tr}\, U^j V^{k-j}, \quad j=0,\ldots,k;
\ \ k=1,\ldots,n,
\end{equation}
\noindent are functionally independent.}

\smallskip

\noindent {\bf Proof.} To prove Lemma 3, it is sufficient to show
that the generic rank of the Jacobi matrix of expressions (1.10)
is equal to $(n(n + 3))/2$ that is the difference between the
number of independent elements of $U$ and $V$ and the rank of the
operators (1.5). We shall limit ourselves to the case when $u_{ab}
= 0$ if $a\not= b$. Then equations (1.10) depend on $(n(n + 3))/2$
variables and their independence is equivalent to the nonvanishing
of the Jacobian.

Let us write down the elements of the Jacobian which are needed
for further reasoning
\begin{equation}
\left|\begin{array}{ccc|ccccccc}
1 & \cdots & 1  & &&&&&&\\
2u_{11} &\cdots & 2 u_{n} & &&& \bf{0} &&&\\
 \cdots&\cdots & \cdots & &&&  &&&\\
nu_{11}^{n-1} &\cdots & nu_{nn}^{n-1} &&&&&&&\\[1mm]
\hline
&& & 1 & 0 & \cdots & 0 & 1 & \cdots & 1\\
& \cdots & & 2v_{11} & 4v_{12} & \cdots & 4v_{1n} & 2v_{22} & \cdots & 2v_{nn}\\
&&& && \cdots &&  &
\end{array}\,\right|.
\end{equation}

Since in the first $n$ rows all the elements besides the first $n$
columns are equal to $j$ zero, the Jacobian (1.11) is equal to the
product of the Jacobian of the elements ${\rm tr}\, U^k$, $k =
1,\ldots,n$, and the Jacobian of all other elements. According to
Lemma~2, the expressions ${\rm tr}\, U^k$, $k = 1,\ldots,n$, are
independent and their Jacobian is not equal to zero; thus, it
remains to show the nonvanishing of the Jacobian and the
functional independence only for the elements
\[
{\rm tr}\, U^j V^{k-j}, \quad j=0,\ldots,k-1; \ \ k=1,\ldots,n.
\]

It follows from (1.11) that it is sufficient to show the
nonvanishing of this Jacobian without the $(n + 1)$th rows and
columns. Thus, to prove the lemma it is enough to show that the
following expressions are independent
\begin{equation}
{\rm tr}\, U^j V^{k-j}V, \quad j=0,\ldots,k;\ \ k=1,\ldots,n-1.
\end{equation}

The above reasoning allows us to make use of the principle of
mathematical induction.

When $n = 1$, $u_{11}$ and $v_{11}$ are independent and the lemma
is true. Let us suppose that it is true for $n - 1$ and then prove
from this that it is valid for $n$. Let the expressions
\begin{equation}
{\rm tr}\, U^j V^{k-j}, \quad j=0,\ldots,k;\ \ k=1,\ldots,n-1,
\end{equation}
\noindent where $U$, $V$ are symmetric $(n-1)\times (n-1)$
matrices and are independent. Then, we shall prove the
independence of (1.12) for the same matrices. The sets (1.12) and
(1.13) coincide with the exception of the following subsets
\begin{equation}
{\rm tr}\, U^j V^{n-j}, \quad j=0,\ldots,n-1
\end{equation}
\noindent belong only to (1.12) and
\begin{equation}
{\rm tr}\, U^j, \quad j=1,\ldots,n-1
\end{equation}
\noindent belong only to (1.13).

The assumption of validity of the lemma for $n-1$ means that for
two symmetric tensors of order~2, the set (1.13) is a functional
basis of invariants of the rotation algebra. Thus, all the
invariants of this algebra can be represented as functions of
(1.13). To prove the functional independence of (1.12), it is
sufficient to prove the nondegeneracy of the Jacobi matrix of the
functions expressing the invariants (1.12) with (1.13). This
matrix has the form
\begin{equation}
\left(\begin{array}{cccc|c|c}
1 & & \bf{0} & & &\\
& 1 & & && \cdots \\
&& \ddots &&&\\
& \bf{0} && 1  &&\\
\hline
&&&& W & \\
\hline &&\bf{0} && &\frac{\partial({\rm
tr}\,U^jV^{n-j})}{\partial({\rm tr}\, U^j)}
\end{array}\right),
\end{equation}
\noindent $W$  being the derivative by ${\rm tr}\, V$ of the
expression
\[
{\rm tr}\, V^n=F({\rm tr}\, V^k,\; k=1,\ldots,n-1).
\]
\noindent (We know that from the Hamilton--Cayley theorem);
$W\not= 0$.

We have only to prove the nonvanishing of the Jacobian of the
expressions
\begin{equation}
{\rm tr}\, (U^j V^{n-j})=F({\rm tr}\,U^k,\;
k=1,\ldots,n-1,\ldots).
\end{equation}
When $V=E$, the corresponding quadrant of the matrix (1.16) is the
unit matrix and its determinant does not vanish identically. This
fact proves the nondegeneracy of the matrix (1.16). The
expressions (1.17) can be obtained from the Hamilton--Cayley
theorem. They are polynomials and, thus, continuous functions of
their arguments.

Functional independence of the expressions (1.12) for $(n-1)\times
(n-1)$ matrices implies their independence for $n\times n$
matrices. From the above, it follows that the expressions (1.10)
are independent, thus Lemma 3 is proved. \hfill\rule{2mm}{2mm}

\smallskip

\noindent {\bf Proof of Theorem 2.} It is easy to see from the
structure of the set (1.3) that the invariants involving
$(u_a^1),\ldots,(u_a^m)$, $(u_{ab}^2),\ldots,(u_{ab}^m)$ depend on
the components of $(u_{ab}^1)$ and of the corresponding vector or
tensor, thus it is sufficient to prove the functions independence
of each of the following sets:
\[
\begin{array}{ll}
R_k(u_a^r, u_{ab}^1) \quad & \mbox{for every} \ r=1,\ldots,m;
\vspace{1mm}\\
S_{jk}(u_{ab}^1,u_{ab}^r) \quad & \mbox{for every} \ r=2,\ldots,m;
\end{array}
\]

Functional independence of each set of $R_k(u_a^r,u_{ab}^1)$ can
be proved similarly to the proof of Lemma~2. The functional
independence of the set $S_{jk}(u_{ab}^1,u_{ab}^r)$ easily follows
from Lemma~3, $u^r$ are evidently independent of other elements of
(1.3).

To make sure that expressions (1.3) are invariants of $AO(n)$, it
is sufficient to substitute them into the condition (1.4).

The set (1.3) consists of
\[
2mn+m+(m-1)\frac{n(n-1)}{2}=m\left(\frac{n(n+1)}{2}+n+1\right)-\frac{n(n-1)}{2}
\]
\noindent elements and, thus, it is complete.

So we have proved that this set forms a basis of invariants for
the algebra $AE(n)$ (1.1).

\smallskip

\noindent {\bf 1.3. Bases of invariants for the extended Euclid
algebra and for the con\-for\-mal algebra.} The extended Euclid
algebra $AE_1(n)$ for one scalar function is defined by the basis
operators $\partial_a$, $J_{ab}$ (1.1) and $D$ depending on a
parameter $\lambda$: \begin{equation} D=x_a\partial_a+\lambda
u\partial_u\quad (\partial_u=\partial/\partial u).
\end{equation}

The basis of the conformal algebra $AC(n)$ consists of the
operators $\partial_a$, $J_{ab}$ (1.1) and $D$ (1.18) and
\begin{equation}
K_a=2x_aD-x_a x_b \partial_a.
\end{equation}

\noindent {\bf Theorem 3.} {\it There is a functional basis for
the extended Euclid algebra that has the following form

(1) when $\lambda\not= 0$:
\begin{equation}
\frac{R_k(u_a,u_{ab})}{u^{k(1-2/\lambda)+1}},\quad
\frac{S_k(u_{ab})}{u^{k(1-2/\lambda)}};
\end{equation}

(2) when $\lambda=0$:
\begin{equation}
u, \quad \frac{R_k(u_a,u_{ab})}{(u_{aa})^k},\quad
\frac{S_k(u_{ab})}{(u_{aa})^k} \quad (k\not=1);
\end{equation}

a functional basis for the conformal algebra has the following
form:

(1) when $\lambda\not=0$:
\begin{equation}
S_k(\theta_{ab})u^{k(2/\lambda-1)};
\end{equation}

(2) when $\lambda=0$:
\begin{equation}
u, \quad S_k(w_{ab})(u_a u_a)^{-2k}\quad (k\not= n),
\end{equation}
\noindent where
\begin{equation}
\begin{array}{l}  \theta_{ab}=\lambda u_{ab}+(1-\lambda)\frac{u_a
u_b}{u}-\delta_{ab}\frac{u_c u_c}{2u},
\vspace{1mm}\\
w_{ab}=u_cu_c\left(u_{ab}+\frac{\delta_{ab}}{2-n}u_{dd}\right)-u_c(u_a
u_{bc}+u_b u_{ac}), \end{array}
\end{equation}
\noindent $\delta_{ab}$ being the Kronecker symbol.}

\smallskip

\noindent {\bf Proof.} To find absolute differential invariants of
the algebra $AE_1(n)$, it is necessary to add to (1.4) the
following condition
\begin{equation}
\mathop{D}\limits^2 F\equiv x_a F_{x_a}+\lambda u
F_u+(\lambda-1)u_a F_{u_a}+ (\lambda-2)u_{ab}F_{u_{ab}}=0.
\end{equation}
Solving equation (1.25) for
\[
F=F(u,R_k(u_a,u_{ab}),S_k(u_{ab})),
\]
\noindent we obtain functional bases (1.20), (1.21) for the
extended Euclid algebra.

The second-order differential invariants of the algebra $AC(n)$
are defined by the conditions (1.4), (1.25) and
\begin{equation}
k_a \mathop{K}\limits^2\!{}_a F=0,
\end{equation}
\noindent where $k_a$ are arbitrary real numbers,
$\mathop{K}\limits^2\!{}_a$ are the second prolongations of the
operators $K_a$ (1.19):
\[
\mathop{K}\limits^2\!{}_a=2x_a \mathop{D}\limits^2 {} +
x_b\mathop{J}\limits^2\!{}_{ab} +2\lambda[u\partial_{u_a}+2u_b
\partial_{u_{ab}}]+2u_a\partial_{u_{cc}}-4u_b \partial_{u_{ab}}.
\]

Solving this system for an arbitrary $n$ requires a lot of
cumbersome computations. It is simpler to construct conformally
covariant tensors from $u$, $u_a$, $u_{ab}$ and then to construct
invariants of the rotation algebra.

\smallskip

\noindent {\bf Definition 3.} {\it Tensors $\theta_a$ and
$\theta_{ab}$ of order 1 and 2 are called covariant with respect
to some algebra $L=\langle J_{ab},X_i\rangle$ if
\begin{equation}
\begin{array}{l}
X_i\theta_a=\sigma_{ab}^i\theta_b +\sigma^i\theta_a,
\vspace{1mm}\\
X_i\theta_{ab}=\rho^i_{ab}\theta_{cb}+\rho^i_{bc}\theta_{ac}+\rho^i\theta_{ab},
\end{array}
\end{equation}
\noindent $X_i$ are operators of the form (0.1), $\rho^i$,
$\sigma^i$ are some functions, $\sigma_{ab}^i$, $\rho_{ab}^i$ are
some skew-symmetric tensors.}

\smallskip

It is easy to show that the expressions $S_k(\theta_{ab})$,
$R_k(\theta_a,\theta_{ab})$, where $\theta_a$, $\theta_{ab}$ are
tensors covariant with respect to the algebra $L$ are relative
invariants of this algebra.

The fact that $\theta_{ab}$ and $w_{ab}$ (1.24) are covariant with
respect to the conformal algebra $AC(n)$ can be verified by direct
substitution of these tensors into the conditions (1.27) for the
operators $\mathop{D}\limits^2$ and $\mathop{K}\limits^2\!{}_a$.

The rank of the second prolongation of the algebra $AC(n)$ is
equal to the number of its operators
\[
\frac{n(n-1)}{2}+n+n+1=\frac{n(n+3)}{2}+1
\]
\noindent and, therefore, a functional basis of second-order
differential invariants must contain $n$ invariants.

Functional independence of the expressions (1.22) follows from
Lemma 2 if we notice that the transformation $u_{ab}\to
\theta_{ab}$ is nondegenerated. The same is true for the set
(1.23).

The expressions (1.22) and (1.23) satisfy (1.25) and (1.26) for
the correspon\-ding~$\lambda$ and they are invariants of the
conformal algebra.

All that is stated above leads to the conclusion that (1.22) and
(1.23) form functional bases for the conformal algebra $AC(n)$
with $\lambda \ne 0$ and $\lambda=0$, respectively.

\smallskip

\noindent {\bf Note 1.} Using condition (1.26), it is easy to show
that when $\lambda\not=0$ covariant tensors exist for $AC(n)$ of
order 2 only; when $\lambda=0$, the tensors $w_{ab}$ (l.24) and
$u_a$ are  conformally covariant but $S_k(w_{ab})$ and
$R_k(u_a,w_{ab})$ are dependent.

\smallskip

\noindent {\bf Theorem 4.} {\it The   second-order   differential
invariants  for a   vector  function $u = (u^1,\ldots, u^m)$ and
for the algebra $AE_1(n) =\langle \partial_a,J_{ab},D\rangle$, the
operator $D$ having the form
\begin{equation}
D=x_a\partial_a+\lambda u^r\partial_{u^r}
\end{equation}
\noindent with a summation over $r$ from 1 to $m$, can be
represented as the functions of the following expressions:

(1) when $\lambda\not=0$:
\[
\frac{u^r}{u^1} \ (r=2,\ldots,m), \quad
\frac{S_{jk}(u_{ab}^1,u_{ab}^r)}{(u^1)^{k(1-2/\lambda)}}, \quad
\frac{R_k(u_a^r,u_{ab}^1)}{(u^1)^{k(1-2/\lambda)+1}};
\]

(2) when $\lambda=0$:
\[
u^r, \quad R_k(u_a^r,u_{ab}^1)(u_{aa}^1)^{-k}, \quad
S_{jk}(u_{ab}^1,u_{ab}^r)(u_{aa}^1)^{-k}
\]
\noindent (when $r=1$ then $k\not=1$);

the    corresponding    basis   for    the    conformal algebra
$AC(n)=\langle \partial_a,J_{ab},D,K_a\rangle$ $(K_a=2x_aD
-x_bx_b\partial_a)$ has the following form:

(1) when $\lambda\not=0$:
\renewcommand{\theequation}{1.29a}
\begin{equation}
\begin{array}{l}
S_{jk}(\theta_{ab}^r,\theta_{ab}^1)(u^1)^{k(2/\lambda-1)}, \quad
\frac{u^r}{u^1},\vspace{1mm}\\
R_k(\theta_a^r,\theta_{ab}^1)^{k(2/\lambda-1)-1}\quad
(r=2,\ldots,m);
\end{array}
\end{equation}

(2) when $\lambda=0$:
\renewcommand{\theequation}{1.29b}
\begin{equation}
\begin{array}{l}
u^r \ (r=1,\ldots,m), \quad
(u_d^1u_d^1)^{-2k}S_{jk}(w_{ab}^1,w_{ab}^r),
\vspace{1mm}\\
(u_d^1u_d^1)^{1-2k} R_k(u_a^r,w_{ab}^1)\quad (r=2,\ldots,m)
\end{array}
\end{equation}
\noindent (for the set of invariants $(u_d^1u_d^1)^{-2k}
S_k(w_{ab})$, $k$ does not take the value $n$); the tensors
$\theta_{ab}^r$, $w_{ab}^r$ are constructed similarly to (1.24)
and}
\[
\theta_a^r=\frac{u_a^r}{u^r}-\frac{u_a^1}{u^1}.
\]

Theorem 4 is proved similarly to Theorem 3.

Functional independence of the sets of invariants follows from
Lemma 2 and 3 taking into account the fact that transformations
$u_{ab}^r\to \theta_{ab}^r$, $u_{ab}^r\to w_{ab}^r$
$(r=1,\ldots,m)$ and $u_a^r\to \theta_a^r$ $(r=2,\ldots,m)$ are
nondegenerate.

\smallskip
\noindent {\bf 1.4. Differential invariants of the rotation
algebra.} The rotation algebra is defined by the basis operators
$J_{ab}$ (1.1).

The second-order invariants of this algebra for $m$ scalar
functions $u^r$ are construc\-ted with $x_a$, $u^r$, $u_a^r$,
$w_{ab}^r$ similarly to invariants of the Euclid algebra.
\smallskip

\noindent {\bf Theorem 5.} {\it  There is a functional basis of
the second-order differential invariants for the algebra $AO(n)$
that has the form
\[
u^r,\quad S_{jk}(u_{ab}^1,u_{ab}^r),\quad
R_k(u_a^r,u_{ab}^1),\quad R_k(x_a,u_{ab}^1),\quad r=1,\ldots,m;
\]
\noindent the corresponding basis of invariants for the algebra
$\langle J_{ab},D\rangle$, where $D$ is defined by (1.28),
consists of the expressions
\[
\begin{array}{l}
\frac{u^r}{u^1} \ (r=2,\ldots,m),\quad
\frac{S_{jk}(u_{ab}^1,u_{ab}^r)}{(u^1)^{k(1-2/\lambda)}}, \quad
R_k(u_a^r,u_{ab}^1)(u^1)^{2k/\lambda-1-k},
\vspace{1mm}\\
R_k(x_a,u_{ab}^1)(u^1)^{2/\lambda(k-2)-k+1}, \quad when \
\lambda\not=0;
\vspace{1mm}\\
u^r,\quad R_k(u_a^r,u_{ab}^1)(u_{aa}^1)^{-k}, \quad
S_{jk}(u_{ab}^1,u_{ab}^r)(u_{aa}^1)^{-k} \ (k\not=1 \ when \ r=1),
\vspace{1mm}\\
R_k(x_a,u_{ab}^1)(u_{aa}^1)^{2-k} \quad when \ \lambda=0.
\end{array}
\]
A basis of invariants for the algebra $\langle
J_{ab},D,K_a\rangle$ when $\lambda\not=0$, consists of the
expres\-sions (1.29a) and
\[
\frac{R_k(x_a,\theta_{ab}^1)}{x^2(u^1)^{(k-1)(1-2/\lambda)}},\quad
k=2,\ldots,n+1;
\]
\noindent when $\lambda=0$ it consists of the expressions (1.29b)
and}
\[
\frac{R_k(x_a,w_{ab}^1)}{x^2(w_{aa}^1)^{k-1}}\quad (x^2=x_ax_a).
\]

The proof of this theorem is similar to the proofs of Theorems 2
and 3; notice that $(x_a)$ is a covariant tensor with respect to
the conformal operators.

\medskip

\renewcommand{\theequation}{2.\arabic{equation}}
\setcounter{equation}{0}

\centerline{\bf 2. Differential invariants of the Poincar\'e and
conformal algebra}

In this section, we consider differential invariants of the second
order for a set of~$m$ scalar functions
\[
u^r=u^r(x_0,x_1,\ldots,x_n),\quad n\geq 3.
\]

The Poincar\'e algebra $AP(1, n)$ is defined by the basis
operators \begin{equation}
p_\mu=ig_{\mu\nu}\frac{\partial}{\partial x_\mu}, \quad
J_{\mu\nu}=x_\mu p_\nu -x_\nu p_\mu, \end{equation} where $\mu$,
$\nu$ take the values $0,1,\ldots,n$; the summation is implied
over the repeated indices (if they are small Greek letters) in the
following way: \begin{equation} x_\nu x^\nu\equiv x_\nu
x^\nu\equiv x^\nu x_\nu=x_0^2-x_1^2-\cdots-x_n^2, \quad
g_{\mu\nu}={\rm diag}\,(1,-1,\ldots,-1). \end{equation}

We consider $x_\nu$ and $x^\nu$ equal with respect to summation,
not to mix signs of derivatives and numbers of functions.

The quasilinear second-order invariants of the Poincar\'e algebra
were described in~[12].

\smallskip

\noindent {\bf Theorem 6.} {\it There is a functional basis of the
second-order differential invariants of the Poincar\'e algebra
$AP(l,n)$ for a set of $m$ scalar functions $u^r$ consisting of
\[
m(2n + 3) + (m - 1)\frac{n(n + 1)}{2}
\]
\noindent invariants
\[
u^r,\quad R_k(u_\mu^r,u_{\mu\nu}^1),\quad
S_{jk}(u_{\mu\nu}^r,u_{\mu\nu}^1).
\]
In this section, everywhere $k = 1,\ldots, n + 1$; $j = 0,\ldots,
k$; $r = 1,\ldots, m$.

For the extended Poincar\'e algebra $A\tilde P(l,n) = \langle
p_\mu,J_{\mu\nu},D\rangle$, where \begin{equation} D=x_\mu p_\mu
+\lambda u^rp_{u^r} \end{equation} $(p_{u^r} = i(\partial/\partial
u^r)$, the summation over $r$ from 1 to $m$ is implied) the
corresponding basis has the following form:

(1) when $\lambda=0$:
\[
u^r,\quad
S_{jk}(u_{\mu\nu}^r,u_{\mu\nu}^1)(u_{\alpha\alpha}^1)^{-k},\quad
R_k(u_\mu^r,u_{\mu\nu}^1)(u_{\alpha\alpha}^1)^{-k};
\]

(2) when $\lambda\not=0$:
\[
\frac{u^r}{u^1},\quad S_{jk}(u_{\mu\nu}^r,u_{\mu\nu}^1)
(u^1)^{k(2/\lambda-1)},\quad
R_k(u_\mu^r,u_{\mu\nu}^1)(u^1)^{2k/\lambda-k-1},
\]
\noindent where $S_{jk}$, $R_k$ are defined similarly to (0.3) and
the summation over small Greek indices is of the type (2.2).

For the conformal algebra $AC(1, n) = \langle
p_\mu,J_{\mu\nu},D,K_\mu\rangle$, where
\[
K_\mu=2x_\mu D-x_\nu x_\nu p_\mu
\]
\noindent ($D$ being the dilation operator (2.3)), the
corresponding basis consists of the expres\-sions
\[
S_{jk}(\theta_{\mu\nu}^r,\theta_{\mu\nu}^1)(u^1)^{k(2/\lambda-1)},\quad
\frac{u^r}{u^1}, \quad
R_k(\theta_\mu^r,\theta_{\mu\nu}^1)(u^1)^{k(2/\lambda-1)-1};
\]
\noindent when $\lambda\not=0$; $r=2,\ldots,m$, there is no
summation over $r$; the conformally covariant tensors have the
form
\[
\theta_\mu^r=\frac{u^r_\mu}{u^r}-\frac{u^1_\mu}{u^1},\quad
\theta_{\mu\nu}^r=\lambda
u_{\mu\nu}^r+(1-\lambda)\frac{u_\mu^ru_\nu^r}{u^r}-
g_{\mu\nu}\frac{u_\beta^r u_\beta^r}{2u^r}.
\]

When $\lambda=0$, the corresponding basis of invariants for the
conformal algebra has the form
\[
u^r,\quad S_{jk}(w_{\mu\nu},w_{\mu\nu}^1)(u^1_\alpha
u^1_\alpha)^{-2k},\quad R_k(u_\mu^r,w_{\mu\nu}^1)(u^1_\alpha
u^1_\alpha)^{1-2k},\quad r=2,\ldots,m;
\]
\noindent the tensors $(w_{\mu\nu}^r)$,
\[
w_{\mu\nu}^r=u_\alpha^r u_\alpha^r
\left(u_{\mu\nu}^r-\frac{g_{\mu\nu}}{1-n}
u_{\beta\beta}^r\right)-u_\beta^r(u_\mu^r u_{\beta\nu}^r +u_\nu^r
u_{\beta\mu}^r)
\]
\noindent are conformally invariant (there is no summation over
$r$).}

The proof of Theorem 6 follows from those of Theorems 2, 3 for
$x=(x_1,\ldots,x_{n+1})$ if we substitute $ix_0$ instead of
$x_{n+1}$.

Similarly to the results of Paragraph 1.4, it is possible to
construct the invariants of the algebras $\langle
J_{\mu\nu}\rangle$, $\langle J_{\mu\nu},D\rangle$, $\langle
J_{\mu\nu},D,K_\mu\rangle$.

The obtained results allow us to construct new nonlinear
many-dimensional equa\-tions, e.g. the equation
\[
\frac{u_\alpha u_\alpha}{1-n}u_{\nu\nu}-u_\mu u_\nu
u_{\mu\nu}=(u_\nu u_\nu)^2 F(u),
\]
\noindent where $F$ is an arbitrary function, is invariant under
the algebra $AC(1, n)$, $\lambda=0$. The left-hand part of the
above equation is equal to $w_{\mu\mu}$.

There is another quasi-linear relativistic equation with rich
symmetry properties
\[
(1-u_\alpha u_\alpha)u_{\mu\mu}-u_\alpha u_\mu u_{\alpha \mu}=0,
\]
\noindent that is, the Born--Infeld equation. The symmetry and
solutions of this equation were investigated in [10, 13]. This
equation is invariant under the algebra $AP(1, n + 1)$ with the
basis operators
\[
J_{AB}=x_A p_B-x_Bp_A,
\]
$A,B=1,\ldots,n+1$, $x_{n+1}\equiv u$.

Let us consider the class of equations
\[
u_{\mu\nu} u_{\mu\nu}=F(u_{\mu\mu},u_\mu u_\nu u_{\mu\nu}, u_\mu
u_\mu,u).
\]
It is evident that they are invariant with respect to the
Poincar\'e algebra $AP(1, n)$ out the straightforward search the
conformally invariant equations from this class with the standard
Lie technique requires a lot of cumbersome calculations. The use
of differential invariants turns this problem into one of
elementary algebra, e.g. if $\lambda\not=0$
\[
F-u_{\mu\nu} u_{\mu\nu}=-\frac{1}{\lambda}S_2(\theta_{\mu\nu})+
u^{2(1-2/\lambda)}\phi(S_1(\theta_{\mu\nu}) u^{2/\lambda-1}),
\]
\noindent where $\theta_{\mu\nu}$ is of the form (1.24) and $\phi$
is an arbitrary function. Whence
\[
\begin{array}{l}  F
=u^{2(1-2/\lambda)}\phi\left(u^{2/\lambda-1}\left(u_{\mu\mu}-\frac{\lambda+n}{\lambda}
\frac{u_\alpha u_\alpha}{u}\right)\right)-
\vspace{2mm}\\
\phantom{F=}{}-\frac{1}{\lambda^2 u^2}(\lambda^2+n^2)(u_\alpha
u_\alpha)^2-\frac{2(1-\lambda)}{\lambda u} u_\mu u_\nu u_{\mu\nu}+
\frac{2u_{\mu\mu} u_\alpha u_\alpha}{\lambda u}. \end{array}
\]

It is useful to note that besides the traces of matrix powers
(0.3), one can utilize all possible invariants of covariant
tensors $\theta_{\mu\nu}^r$, $w_{\mu\nu}^r$ to construct
conformally invariant equations.

\medskip

\renewcommand{\theequation}{3.\arabic{equation}}
\setcounter{equation}{0}

\centerline{\bf 3. Differential invariants of an
infinite-dimensional algebra}

It is well-known that the simplest first-order relativistic
equation --- the eikonal or Hamilton equation
\begin{equation}
u_\alpha u_\alpha\equiv u_0^2-u_1^2-\cdots-u_n^2=0
\end{equation}
\noindent is invariant under the infinite-dimensional algebra
$AP^\infty(1,n)$ generated by the opera\-tors [10, 14]
\begin{equation}
X=(b^{\mu\nu} x_\nu +a^\mu)\partial_\mu+\eta(u)\partial_u,
\end{equation}
\noindent $-b^{\mu\nu}=b^{\nu\mu}$, $a^\mu$, $\eta$ being
arbitrary differentiate functions on $u$. Equation (3.1) is widely
used in geometrical optics.

In this section, we describe a class of second-order equations
invariant under the algebra (3.2).

It is easy to show that the tensor of the rank 2
\begin{equation}
\theta_{\mu\nu}=u_\mu u_{\lambda\nu} u_\lambda+u_\nu
u_{\lambda\mu} u_\lambda-u_\mu u_\nu u_{\lambda\lambda}-u_\lambda
u_\lambda u_{\mu\nu}
\end{equation}
\noindent is covariant under the algebra $AP^\infty(1,n)$ (3.2).

\smallskip

\noindent {\bf Theorem 7.} {\it The equations of the form
\begin{equation}
S_k(\theta_{\mu\nu})=0, \quad k=1,2,\ldots,
\end{equation}
\noindent $S_k$ being defined as (0.3), are invariant with respect
to the algebra $AP^\infty(1,n)$ (3.2).}

\smallskip

The problem of the description of all such equations is more
difficult and we do not consider it here.

Let us investigate in more detail the quasi-linear second-order
equation of the form
\begin{equation}
u_\mu u_{\mu\nu} u_\nu -u_\mu u_\mu u_{\alpha\alpha}=0.
\end{equation}

\noindent {\bf Theorem 8.} {\it When $n\geq 2$, equation (3.5) is
invariant with respect to the algebra $A\tilde P^\infty(1,n)$ with
generators of the form
\[
X+d(u)x_\mu\partial_\mu,
\]
\noindent $X$ is of the form (3.2), $d(u)$ is an arbitrary
function on $u$.}

\smallskip

The proofs of Theorems 7 and 8 can be easily obtained with the Lie
technique using the criterion of invariance
\[
\mathop{X}\limits^2
S_k(\theta_{\mu\nu})\Big|_{S_k(\theta_{\mu\nu})=0}=0,
\]
\noindent where $\mathop{X}\limits^2$ is the second prolongation
of the operator $X$ [8--10].

\medskip

\renewcommand{\theequation}{4.\arabic{equation}}
\setcounter{equation}{0}

\centerline{\bf 4. Differential invariants of the Galilei algebra}

{\bf 4.1.} It is well-known that the heat equation
\begin{equation}
\begin{array}{l}
2\mu u_t+\Delta u=0,\quad \Delta u\equiv u_{aa},
\vspace{1mm}\\
u=u(t,\bf{x}),\quad \bf{x}=(x_1,\ldots,x_n),\quad n\geq 3
\end{array}
\end{equation} \noindent is invariant under the generalized Galilei algebra
$AG^I_2(1,n)$ with the basis operators \begin{equation}
\begin{array}{l}
\partial_t=\frac{\partial}{\partial t},\quad
\partial_a=\frac{\partial}{\partial x_a},\quad J_{ab}=x_a\partial_b-x_b\partial_a,
\vspace{1mm}\\
G_a=t\partial_a+\mu x_au\partial_u \quad
\left(\partial_u=\frac{\partial}{\partial u}\right), \quad
u\partial_u, \quad D=2t\partial_t+x_a\partial_a+\lambda
u\partial_u,
\vspace{1mm}\\
A=tD-t^2\partial_t+\frac{\mu{\partial{x}}^2}{2}u\partial_u\quad
\left(\lambda=-\frac n2\right). \end{array}
\end{equation}

The Schr\"odinger equation \begin{equation} 2im\psi_t+\psi_{aa}=0,
\end{equation} $\psi=\psi(t,\bf{x})$ being a complex-valued
function, is also invariant  [16]  under the gene\-ra\-li\-zed
Galilei algebra with the basis operators \begin{equation}
\begin{array}{l}  p_0=i\frac{\partial}{\partial
t},\quad p_a=-i\frac{\partial}{\partial x_a}, \quad
J_{ab}=x_ap_b-x_bp_a,\quad
J=i(\psi\partial_\psi-\psi^*\partial_{\psi^*}),
\vspace{1mm}\\
G_a=tp_a-mx_a J, \quad D=2tp_0-x_ap_a+\lambda I\quad
(I=\psi\partial_\psi+\psi^*\partial_{\psi^*}),
\vspace{1mm}\\
A=t^2p_0-tx_ap_a+\lambda tI+\frac{m{\ bf{x}}^2}{2}J\quad
\left(\lambda=-\frac n2\right). \end{array}
\end{equation}
The asterisk means the complex conjugation.

We shall designate the algebra (4.4) with the symbol $AG^{II}_2(1,
n)$. Besides,
\[
AG^I(1,n)=\langle
\partial_t,\partial_a,u\partial_u,G_a,J_{ab}\rangle,
\]
\noindent the operators being of the form (4.2). A basis of the
algebra $AG^I_1(1,n)$ consists of the basis operators or
$AG^I(1,n)$ and of the operator $D$. Furthermore
$AG^{II}(1,n)=\langle p_0,p_a,J,J_{ab},G_a\rangle$ (4.4). A basis
of the algebra $AG_1^{II}(1,n)$ consists of the previous operators
and also $D$ (4.4).

To simplify the form of invariants, we introduce the following
change of dependent variables:
\begin{equation}
u=\exp\varphi,\quad \psi=\exp\phi\quad \left({\rm
Im}\phi=\arctan\frac{{\rm Im}\psi}{{\rm Re}\psi}\right).
\end{equation}

All the indices $k$ in the expressions of the type (0.3) here will
take on values from 1 to $n$, the indices $j$ will take on values
from 0 to $k$.

We seek invariants of the algebra $AG^I_2(1,n)$ in the form
\begin{equation}
F=F(\varphi_t,\varphi_a,\varphi_{tt},\varphi_{at},\varphi_{ab}).
\end{equation}
Obviously, they do not include $\varphi$, $x_a$, and $t$ because
the basis (4.2) contains operators $\partial_\varphi$,
$\partial_a$, $\partial_t$.

Using the definition of an absolute differential invariant (0.2)
we get the following conditions on the function $F$ (4.6):
\begin{equation}
\mathop{J}\limits^2\!{}_{ab}F=\varphi_a F_{\varphi_b}-\varphi_b
F_{\varphi_a}+
F_{\varphi_{bt}}\varphi_{at}-\varphi_{bt}F_{\varphi_{at}}+2\varphi_{ac}F_{\varphi_{bc}}-
2\varphi_{bc}F_{\varphi_{ac}}=0,
\end{equation}
\begin{equation}
\mathop{G}\limits^2\!{}_{a}F=-\varphi_aF_{\varphi_t}+\mu
F_{\varphi_a}-2\varphi_{at}
F_{\varphi_{tt}}-\varphi_{ab}F_{\varphi_{bt}}=0,
\end{equation}
\begin{equation}
\mathop{D}\limits^2 F= -2\varphi_t
F_{\varphi_t}-\varphi_aF_{\varphi_a}-4\varphi_{tt}
F_{\varphi_{tt}}-3\varphi_{at}F_{\varphi_{at}}-2\varphi_{ab}F_{\varphi_{ab}}=0,
\end{equation}
\begin{equation}
\mathop{A}\limits^2 F=t \mathop{D}\limits^2 F +x_a
\mathop{G}\limits^2\!{}_{a}F- \lambda F_{\varphi_t}-2\varphi_t
F_{\varphi_{tt}}-\varphi_aF_{\varphi_{at}}+\mu
\delta_{ab}F_{\varphi_{ab}}=0.
\end{equation}

From equations (4.8), we can see that the tensors
\begin{equation}
\theta_a=\mu\varphi_{at}+\varphi_b\varphi_{ab}, \quad \varphi_{ab}
\end{equation}
\noindent are covariant with respect to the algebra $AG^I(1,n)$
$(\mu\not=0)$.

\smallskip

\noindent {\bf Theorem 9.} {\it There is a functional basis of
absolute differential invariants for the algebra $AG^I(1,n)$, when
$\mu\not=0$, consisting of these $2n + 2$ invariants:
\begin{equation}
\begin{array}{l}
M_1=2\mu\varphi_t+\varphi_a\varphi_a,\quad
M_2=\mu^2\varphi_{tt}+2\mu\varphi_a
\varphi_{at}+\varphi_a\varphi_b\varphi_{ab},
\vspace{1mm}\\
R_k=R_k(\theta_a,\theta_{ab}),\quad S_k=S_k(\varphi_{ab}).
\end{array}
\end{equation}
For the algebra $AG^I_1(1,n)$ $(\mu\not=0)$ such a basis has the
form
\begin{equation}
\frac{M_2}{M_1^2},\quad \frac{R_k}{M_1^{2+k}},\quad
\frac{S_k}{M_1^k}.
\end{equation}
For the algebra $AG^I_2(1,n)$ $(\mu\not=0)$, there is a basis of
the form
\begin{equation}
\frac{N_2}{N_1^2},\quad \frac{\hat R_k}{N_1^{2+k}},\quad
\frac{\hat S_k}{N_1^k} \quad (k=2,\ldots,n),
\end{equation}
\noindent where
\begin{equation}
\begin{array}{l}
N_1=2\mu\varphi_t+\varphi_a\varphi_a+\varphi_{aa},
\vspace{1mm}\\
N_2=\mu^2\varphi_{tt}+2\mu\left(\frac
1n\varphi_t\varphi_{aa}+\varphi_a\varphi_{at}
\right)+\varphi_a\varphi_b\varphi_{ab}+\frac
1n\varphi_a\varphi_a\varphi_{bb}+ \frac 1n\varphi_{bb}^2,
\vspace{1mm}\\
\hat R_k=\sum_{l=0}^k
R_l(\varphi_{aa})^{k-1}\frac{(-n)^lk!}{l!(k-l)!},
\vspace{1mm}\\
\hat S_k=\sum_{l=0}^k\frac{(-n)^l(k-1)!(k+1)}{(l+1)!(k-l)!}
S_l(\varphi_{aa})^{k-l},
\end{array}
\end{equation}
\noindent $S_k$, $R_k$ are defined by (4.12) and $\theta_a$ has
the form (4.11).}

\smallskip

The proof of this theorem is similar to the proof of Theorems 2
and 3. We shall present here only some hints to the proof.

It is evident that the function $F$ must depend on the invariants
of the Euclid algebra
\[
F=F(\varphi_t,\varphi_{tt},R_k(\varphi_a,\varphi_{ab}),R_k(\varphi_{at},
\varphi_{ab}),S_k{(\varphi_{ab}})).
\]
First we construct two invariants of $AG^I(1,n)$ $M_1$ and $M_2$
(4.12) which depend on $\varphi_t$ and $\varphi_{tt}$
respectively. The other invariants of the adduced basis (4.12) do
not depend on $\varphi_t$ or $\varphi_{tt}$ and the sets
$\{M_1,M_2\}$ and $\{R_k,S_k\}$ are independent. The invariants
$R_k$, $S_k$ are constructed with the covariant tensors
$\theta_a$, $\varphi_{ab}$ (4.11) similarly to invariants of the
conformal algebra investigated above, and it is easy to see that
they are independent.

The generic ranks of the prolonged algebras $AG^I(1,n)$,
$AG_1^I(1,n)$, $AG_2^I(1,n)$ are equal to the numbers of their
operators and from this fact we can compute the number of elements
in the bases for these algebras.

Adding to (4.7) and (4.8) the condition (4.9), we obtain from the
invariants (4.12) the basis (4.13) for the algebra $AG^I_1(1,n)$.

Relative invariants $\hat R_k$, $\hat S_k$ (4.15) of the algebra
$AG_2^I(1,n)$ were found from the equation
\[
\lambda F_{\varphi_t}-2\varphi_t F_{\varphi_{tt}}-\varphi_a
F_{\varphi_{at}} +\mu\delta_{ab}F_{\varphi_{ab}}=0,
\]
\noindent $F=F(R_k,S_k)$, and then we constructed absolute
invariants using (4.9). Besides, it is possible to construct
analogues to $\hat R_k$, $\hat S_k$ with $AG^I_2(1,n)$-covariant
tensors $\theta_a$ (4.11) and
\[
\theta_{ab}=\varphi_{ab}-\frac{2\delta_{ab}}{n}(\varphi_c\varphi_c
+\mu\varphi_t).
\]

Considering $(\varphi_{at})$, $(\varphi_a)$, $(\varphi_{ab})$
 as independent vectors and tensors and putting
$\varphi_{ab}=0$ whenever $a\not= b$, $\varphi_a=0$, we see from
Lemma 2 that the adduced sets of invariants are independent.

\smallskip

\noindent {\bf Note 2.} A basis of invariants for the Galilei
algebra without translations contains expressions (4.12) and
\[
R_k(h_a,\phi_{ab}),\quad \frac 12\mu{\bf{x}}^2-\varphi t,
\]
\noindent the Galilei-covariant vector $h_a$ having the form
\[
h_a=\mu x_a -t\varphi_a.
\]

Let us also adduce an $A$-covariant tensor
\[
\hat h_a=\frac{\mu x_a}{t}-\varphi_a
\]
\noindent depending on $x_a$, and a relative invariant of the
operators $A$ and $D$ (4.2)
\[
\exp\left\{\varphi-\frac{\mu{\bf{x}}^2}{2t}\right\}
\]
\noindent with which it is possible to construct a basis of
invariants for the algebra $\langle G_a,J_{ab},D$, $A\rangle$.

We have presented a method to find the bases of invariants for Lie
algebras for which $J_{ab}$ (1.1) are basis operators. Further, we
shall adduce functional bases for the algebras $AG^I_2(1,n)$ where
$\mu=0$
 and $AG^{II}_2(1,n)$ where $\mu=0$ or $\mu \ne 0$. We omit proofs
because they are similar to proofs of the previous theorems.

It is evident from the conditions (4.7)--(4.10) that the case
$\mu=0$
 for the algebra $AG^I_2(1,n)$ has to be specially considered. The
tensors $(\varphi_a)$ and $(\varphi_{ab})$ are covariant with
respect to this algebra; the tensor $(\theta_a)$ involved in
invariants is defined by an implicit correlation
\begin{equation}
\varphi_{bt}=\theta_a\varphi_{ab}.
\end{equation}

\noindent {\bf Theorem 10.} {\it There is a functional basis of
the second-order differential invariants for the algebra
$AG^I(1,n)$, where $\mu=0$, that has the form
\begin{equation}
\begin{array}{l}
M_1=\varphi_t-\varphi_a\theta_a,\quad
M_2=\varphi_{tt}-\varphi_{at}\theta_a, \vspace{1mm}\\
R_k=R_k(\varphi_a,\varphi_{ab}),\quad S_k=S_k(\varphi_{ab}).
\end{array}
\end{equation}

The corresponding basis for the algebra $AG_1^I(1,n)$, where
$\mu=0$ has the form
\[
\frac{M_1^2}{M_2},\quad \frac{R_k}{M_1^k},\quad \frac{S_k}{M_1^k};
\]
for the algebra $AG_2^I(1,n)$, when $\mu=0$, it has the form
\[
\frac{R_k}{M^{1/2k}},\quad \frac{S_k}{M^{1/2k}},
\]
\noindent where $R_k$, $S_k$ are defined by (4.17) and
\[
M=(\varphi_t-\theta_a\varphi_a)^2+(\varphi_{tt}-\varphi_{at}\theta_a)
(\lambda+\varphi_a\varphi_{b}r_{ab}).
\]
Here, the matrix $\{r_{ab}\} = \{\varphi_{ab}\}^{-1}$;
$\theta_a=r_{ab}\varphi_{bt}$ are the same as in (4.16).}

\noindent {\bf Note 3.} It is possible to use, instead of $M_1$,
$M_2$, the invariants
\[
\hat M_1=\left|\begin{array}{cccc}
\varphi_t &\varphi_1 & \cdots &\varphi_n\\
\varphi_{1t} & \varphi_{11} & \cdots &\varphi_{1n}\\
\cdots & \cdots & \cdots &\cdots\\
\varphi_{nt} & \varphi_{n1} &\cdots &
\varphi_{nn}\end{array}\right|,\qquad \hat
M_2=\left|\begin{array}{cccc}
\varphi_{tt} &\varphi_{1t} & \cdots &\varphi_{nt}\\
\varphi_{1t} & \varphi_{11} & \cdots &\varphi_{1n}\\
\cdots & \cdots & \cdots &\cdots\\
\varphi_{nt} & \varphi_{n1} &\cdots &
\varphi_{nn}\end{array}\right|,
\]
\noindent which have been found in [17] as the solution of the
problem of finding the equations invariant under the Galilei
algebra when $\mu=0$.

\smallskip

\noindent {\bf Note 4.} The invariants for the algebra $\langle
J_{ab},G_a,J,D,A\rangle$ (4.2), where $\mu=0$, which depend on
$x_a$, $t$, can be constructed with $\varphi_a$, $\varphi_{ab}$
and the following covariant vector
\[
\hat h_a=\frac{h_a}{t}+\frac 2nt\varphi_a\varphi_t+\frac
4n\frac{x_b\varphi_b\varphi_a}{t},
\]
\noindent where $h_a=x_b\varphi_{ab}+t\varphi_{at}$ is covariant
with respect to the operators $G_a$ when $\mu=0$.

\smallskip
\noindent {\bf 4.2.} Let us proceed to describe the basis of the
invariants for the algebra $AG^{II}_2(1,n).$

\smallskip

\noindent {\bf Theorem 11.} {\it Any absolute differential
invariant of order $\leq 2$ for the algebras listed below is a
function of the following expressions:

(1) $AG^{II}(1,n)$, $m\not=0$:
\[
\begin{array}{l} \phi+\phi^*,\quad M_1=2im\phi_t+\phi_a\phi_a,\quad M_1^*,
\vspace{1mm}\\
M_2=-m^2\phi_{tt}+2im\phi_a\phi_{at}+\phi_a\phi_b\phi_{ab},\quad
M_2^*,
\vspace{1mm}\\
S_{jk}=S_{jk}(\phi_{ab},\phi_{ab}^*),\quad
R_k^1=R_k(\theta_a,\phi_{ab}),
\vspace{1mm}\\
R_k^2=R_k(\theta^*_a,\phi_{ab}),\quad
R_k^3=R_k(\phi_a+\phi^*_a,\phi_{ab}), \end{array}
\]
\noindent
the covariant tensors being
$\theta_a=-im\phi_{at}+\phi_b\phi_{ab}$;

(2) $AG^{II}_1(1,n)$, $m\not=0$:
\[
\frac{M_1^*}{M_1},\quad \frac{M_2}{M_1^2},\quad
\frac{M_2^*}{M_1^2}, \quad \frac{R_k^l}{M_1^{2+k}} \ (l=1,2),
\quad \frac{R_k^3}{M_1^k},\quad \frac{S_{jk}}{M_1^k},
\]
\noindent $\phi+\phi^*$ \ when \ $\lambda=0$, $
M_1e^{(2/\lambda)(\phi+\phi^*)} $\ \ when \ $\lambda\not=0$.

\smallskip
(3) $AG_2^{II}(1,n)$, $m\not=0$, $\lambda=-\frac n2$:
\[
N_1 e^{(-4/n)(\phi+\phi^*)},\quad \frac{N_1}{N_1^*},\quad
\frac{N_2}{N_1^2},\quad \frac{N_2^*}{N_1^2},\quad \frac{\hat
R_k^l}{N_1^{2+k}} \ (l=1,2),\quad \frac{\hat R_k^3}{N_1^k},\quad
\frac{\hat S_{jk}}{N_1^k},
\]
\noindent where
\[
\begin{array}{l} N_1=2im\phi_t+\phi_{aa}+\phi_a\phi_a,
\vspace{1mm}\\
N_2=-m^2\phi_{tt}+2im\left(\phi_a\phi_{at}+\frac
1n\phi_t\phi_{aa}\right)+ \phi_a\phi_b\phi_{ab}+\frac
1n\phi_a\phi_a\phi_{bb}+\frac 1n\phi_{aa}^2,
\vspace{1mm}\\
 \hat S_{jk}=\sum_{l=0}^k\sum_{r=0}^j S_{rl} (-n)^l C_j^r
C_k^{l+1-r}(\phi_{aa})^{j-r}
(\phi_{aa}^*)^{k-l-j+r}+k(\phi_{aa})^j(\phi_{aa}^*)^{k-j-1},
\vspace{1mm}\\
 \hat R_k^l=\sum_{j=0}^k R_j^l(\phi_{aa})^{k-j}\frac{(-n)^j
k!}{j!(k-j)!} \ \ (l=1,2,3). \end{array}
\]

The invariants for the algebras $AG^{II}(1,n)$, $AG_1^{II}(1,n)$
$(m=0)$ can be con\-st\-ruc\-ted similarly to the case of real
function. Let us adduce a functional basis for the algebra
$AG_2^{II}(1,n)$.

(1) when $\lambda=0$, then there is a basis consisting of the
following expressions:
\[
\phi+\phi^*,\quad \frac{N_1^2}{N_2^2},\quad
\frac{N_1^{*2}}{N_2},\quad \frac{(S_{jk})^2}{N_1^k},\quad
(R_k^l)^2N_1^{-k-1} \ (l=1,2,4);
\]

(2) $\lambda\not=0$:
\[
N_1e^{(4/\lambda)(\phi+\phi^*)},\quad\frac{N_1^*}{N_1},\quad N_3
e^{(3/\lambda)(\phi+\phi^*)},\quad \frac{(R_k^l)^2}{N_1^k} \
(l=1,2,3),\quad \frac{(S_{jk})^2}{N_1^k}, \
\]
\noindent where
\[
\begin{array}{l}
N_1=(\phi_t-\theta_a\phi_a)^2+(\phi_{tt}-\theta_a\phi_{at})(\lambda+\phi_a\phi_{ab}r_{ab})
\vspace{1mm}\\
\qquad (with \ \{r_{ab}\}=\{\phi_{ab}\}^{-1} \ and \
\theta_a=r_{ab}\phi_{bt}),
\vspace{1mm}\\
N_2=(\phi_t-\phi_c\theta_c)\phi_a^*\phi_b^*r_{ab}^*-(\phi_t^*-\phi_c^*\theta_c^*)\phi_a\phi_br_{ab},
\vspace{1mm}\\
N_3=(\phi_t-\phi_t^*)-\tau_a(\phi_a-\phi_a^*) \ \
(\tau_a(\lambda\phi_{ab}+\phi_a\phi_b)=\phi_b\phi_t+\lambda\phi_{bt}),
\vspace{1mm}\\
R_k^1=R_k(\phi_a,\phi_{ab}),\quad
R_k^2=R_k(\phi_a^*,\phi_{ab}),\quad
R_k^3=R_k(\theta_a-\theta_a^*,\phi_{ab}),
\vspace{1mm}\\
R_k^4=R_k(\rho_a,\phi_{ab}) \ \
(\rho_a=(\phi_t-\theta_b\phi_b)(\phi_c^*r_{ac}-\phi_cr_{ac}^*)-\phi_b\phi_d
r_{bd} (\theta_a-\theta_a^*)). \end{array}
\]}

The proof of this theorem will be easier if we notice that by
putting $\mu=im$ in~(4.4), we obtain operators similar to the
operators (4.2).

The change of variables (4.5) in the adduced invariants allows us
to obtain bases for the algebras $AG_2^I$ and $AG_2^{II}$ in the
representations (4.2) and (4.4). These results can also be
generalized for the case of several scalar functions.

\smallskip
\noindent {\bf 4.3.} Let us present some examples of new invariant
equations
\begin{equation}
\begin{array}{l}  \phi_{tt}+\frac{1}{\mu^2}\left\{ 2\mu\left(\frac
1n\phi_t\phi_{aa}+\phi_a\phi_t\right) +\phi_a\phi_b\phi_{ab}+\frac
1n\phi_a\phi_a\phi_{bb}+\frac 1n\phi_{bb}^2\right\}=
\vspace{1mm}\\
\qquad =(2\mu\phi_t+\phi_a\phi_a+\phi_{aa})^2F, \end{array}
\end{equation}
\begin{equation}
\begin{array}{l}  -m^2\phi_{tt}+2im\left(\phi_a\phi_{at}+\frac
1n\phi_t\phi_{aa}\right)+\phi_a\phi_b\phi_{ab} +\frac
1n\phi_a\phi_a\phi_{bb}+\frac 1n\phi^2_{aa}=
\vspace{1mm}\\
 \qquad =(2im\phi_t+\phi_a\phi_a+\phi_{aa})^2F. \end{array}
\end{equation}

Equations (4.18) and (4.19) are invariant, respectively, under the
algebras \linebreak $AG_2^I(1,n)$, $\mu\not=0$ (4.2), and
$AG_2^{II}(1,n)$, $m\not=0$ (4.4). The $F$'s are arbitrary
functions of the invariants for corresponding algebras.

Evidently, wide classes of invariant equations can be constructed
with the adduced invariants.

\medskip

\centerline{\bf 5. Conclusion}

It is well-known that a mathematical model of physical or some
other phenomena must obey one of the relativity principles of
Galilei or Poincar\'e. Speaking the language of mathematics, it
means that the equations of the model must be invariant under the
Galilei or the Poincar\'e groups. Having bases of differential
invariants for these groups (or for the corresponding algebras),
we can describe all the invariant scalar equations, or sort the
invariant ones out of a set of equations.

The construction of differential invariants for vector and spinor
fields presents more complicated problems. The first-order
invariants for a four-dimensional vector potential had been found
in [18]. The cases of spinor and many-dimensional vector
Poincar\'e-invariant equations and corresponding bases of
invariants are still to be investigated.

\smallskip

\noindent {\bf Note 5.} After having prepared the present paper,
we became acquainted with the article [19] where realizations of
the Poincar\'e group $P(1,1)$ and the corresponding conformal
group were investigated, and all second-order scalar differential
equations invariant under these groups were obtained. Reference
[19] contains bases of absolute differential invariants of the
order 2 for the Poincar\'e, the similitude, and the conformal
groups in $(1+1)$-dimensional Minkowski space for various
realizations of the cor\-res\-pon\-ding Lie algebras.

\smallskip

\noindent {\bf Note 6.} It was noticed by the referee that an
essential misunderstanding arose in the calculation of second
prolongations for differential operators, e.g. in formulae (1.5)
and (1.25).

When we calculate such prolongations with the usual Lie technique
(see, e.g.,~[8]), we imply that action of an operator of the form
$X^{ab}\partial_{u_{ab}}$, where $X^{ab}$ are some functions, is
as follows
\[
X^{ab}\partial_{u_{ab}}(u_{cd} u_{cd})=2X^{ab} u_{ab}, \quad
\partial_{u_{ab}} u_{cd}=\delta_{ac}\delta_{bd}.
\]
With this assumption, $\partial_{u_{ab}} u_{ba}=0$, $a\not=b$.

Otherwise, the second prolongation of the operator $J_{ab}$ (1.1)
will be of the form
\[
\begin{array}{l}
\mathop{J}\limits^2\!\!{}_{ab}=J_{ab}+\hat J_{ab},
\vspace{1mm}\\
\hat
J_{ab}=u_a\partial_{u_b}-u_b\partial_{u_a}+u_{ac}\partial_{u_{bc}}-u_{bc}\partial_{u_{ac}}+u_{ab}
(\partial_{u_{bb}}-\partial_{u_{aa}}). \end{array}
\]

\noindent {\bf Note 7.} The equations which are conditionally
invariant with respect to the Poincar\'e and Galilei algebras were
investigated in [20, 21].

\smallskip

{\bf Acknowledgement.} Authors would like to thank the referees
for valuable com\-ments.

\bigskip
\footnotesize{

}

\begin{thebibliography}{99}
\bibitem{Lie Diff.Inv.}  {Lie S., {\it Math. Ann.}, 1884, {\bf 24}, 52--89.}

\bibitem{Tresse}  {Tresse A., {\it Acta Math.}, 1894, {\bf 18}, 1--88.}

\bibitem{Vessiot}  {Vessiot E., {\it Acta Math.}, 1904, {\bf 28}, 307--349.}

\bibitem{Michal}  {Michal A.D., {\it Proc. Nat. Acad. Sci.}, 1951, {\bf 37}, 623--627.}

\bibitem{FYe Galilei inv}  {Fushchych W.I., Yegorchenko I.A., {\it Dokl AN Ukr. SSR, Ser.
A}, 1989, ¹~4, 29--32.}

\bibitem{FYe Poinc inv}  {Fuschchych W.I., Yegorchenko I.A., {\it Dokl. AN Ukr. SSR,
Ser. A}, 1989, ¹~5, 21--22.}

\bibitem{Spencer}  {Spencer A.J.M., Theory of invariants,  New
York, London, Academic Press, 1971.}

\bibitem{Ovs-eng}  {Ovsyannikov L.V., Group analysis of differential equations, New York,
Academic Press, 1982.}

\bibitem{Olver Book1}  {Olver P., Application of Lie groups to differential
equations, New York, Springer-Verlag, 1987.}

\bibitem{FSS}  {Fushchych W.I., Shtelen W.M., Serov N.I., Symmetry
analysis and exact solutions of nonlinear equations of
mathematical physics, Kiev, Naukova Dumka, 1989 (in Russian);
English version to be published by Kluwer Publishers, 1993.}

\bibitem{Bluman Kumei}  {Bluman G.W., Kumei S., Symmetries and differential
equations, New York, Springer Verlag, 1989.}

\bibitem{F Ye DAN}  {Fuschchych W.I., Yegorchenko I.A.,
{\it Dokl. AN SSSR}, 1988, {\bf 298}, 347--351.}

\bibitem{F Serov DAN}  {Fuschchych W.I., Serov N.I., {\it Dokl. AN SSSR}, 1984, {\bf 278},
847.}

\bibitem{F Sht Nuovo Cim.}  {Fushchych W.I., Shtelen W.M., {\it Lett. Nuovo Cimento}, 1982, {\bf 34}, 498.
}

\bibitem{Goff}  {Goff J.A., {\it Amer. J. Math.}, 1927, {\bf 49}, 117--122.}

\bibitem{Niederer}  {Niederer U., {\it Helv. Phys. Acta}, 1972, {\bf 45}, 802--810.}

\bibitem{F Ch JPA}  {Fushchych W.I., Cherniha R.M., {\it J. Phys. A}, 1985, {\bf 18},
3491--3503.}

\bibitem{Ye preprint complex fields}  {Yegorchenko I.A., Symmetry properties of nonlinear
equations for complex vector fields, Preprint 89.48, Institute of
Mathematics of the Ukr. Acad. Sci, 1989.}

\bibitem{Rideau Winternitz}  {Rideau G., Winternitz P., {\it J. Math. Phys.}, 1990, {\bf 31},
1095--1105.}

\bibitem{F Nik 87}  {Fushchych W.I., Nikitin A.G., Symmetries of Maxwell's
equations, Dordrecht, D. Reidel, 1987.}

\bibitem{F UMZh91}  {Fushchych W.I., {\it Ukrain. Mat. Zh.}, 1991, {\bf 43}, 1456.}

\end{thebibliography}
\end{document}